\documentclass[pra,aps,twocolumn]{revtex4-1}

\usepackage{amssymb}

\usepackage{graphicx}
\usepackage{fontenc}
\usepackage{newlfont}
\usepackage{amssymb}
\usepackage{amsfonts}
\usepackage{wasysym}
\usepackage{graphicx}
\usepackage{epsfig}
\usepackage{amsthm}
\usepackage{color}           
\usepackage{soul}            
\usepackage[colorlinks=true, citecolor=blue, urlcolor=blue ]{hyperref}
\usepackage[normalem]{ulem}
\usepackage{cancel}
\newcommand{\etal}{\textit{et al.}}

\begin{document}

\title[Multipartite Entanglement Trends in  Gapless-gapped Transitions]{Genuine Multipartite Entanglement Trends in  Gapless-gapped Transitions of Quantum Spin Systems}

\author{Anindya Biswas, R. Prabhu, Aditi Sen(De) and Ujjwal Sen}

\affiliation{Harish-Chandra Research Institute, Chhatnag Road, Jhunsi, Allahabad 211 019, India}

\begin{abstract}
We investigate the behavior of genuine multiparticle entanglement, as quantified by the generalized geometric measure, in gapless-to-gapped  
quantum  transitions of one- and two-dimensional quantum spin models.
The investigations are performed in the exactly solvable one-dimensional \(XY\) models, as well as 
two-dimensional frustrated \(J_1-J_2\) models, including the Shastry-Sutherland model. 
The generalized geometric measure shows  non-monotonic features near such transitions in 
the frustrated quantum systems. We also compare the features of the generalized geometric measure
near the quantum critical points with the same for measures
of bipartite quantum correlations. The multipartite quantum correlation measure turns out to be a better indicator 
of quantum critical points than the 
bipartite measures, especially for two-dimensional models.
\end{abstract}


\maketitle

\section{Introduction}

\noindent Recent developments at the interface of quantum information science
and many-body physics 
indicate that quantum correlations, in particular, quantum entanglement~\cite{Rhorodecki} can potentially be a 
universal physical characteristic to investigate many-body phenomena. For example, entanglement
has been proposed as a detector of quantum phase transitions (QPT) in spin systems~\cite{Ssachdev,Mlewenstein,Lamico} and
as a tool to develop and simplify efficient numerical simulations like density matrix renormalization group~\cite{Fverstraete}. 
Of late, the developments involving the numerical simulations using matrix product states~\cite{Aklumper}, 
projected entangled pair states~\cite{Fverstraete2}, and tensor network states~\cite{Hniggerman}, 
have established a strong connection between entanglement and many-body theory.
At the same time, the behavior of entanglement has been investigated in many-body systems like cold atomic gases in 
optical lattices~\cite{Strotzky}, and trapped gaseous Bose-Einstein condensates~\cite{Tzibold}.
On the other hand, information-theoretic quantum correlation measures like quantum discord~\cite{Lhenderson}
have also been used to study critical phenomena in many-body systems~\cite{Rdillenschneider}. In this paper, we focus our 
attention on multisite entanglement of many-body physical systems, integrable and non-integrable.

Characterization of entanglement in physical systems is mostly restricted to bipartite entanglement due to the general
unavailability of computatable measures in the multipartite scenario. However, in some cases, it turns
out that the bipartite entanglement measures can not capture the co-operative phenomena in the system~\cite{Mfyang,Xfqian}, 
and therefore, it is natural to look out for multipartite entanglement measures to investigate such many-body systems~\cite{Mnbera}.

Multipartite entanglement measures, e.g. the geometric measure~\cite{Ashimony} 
(cf.~\cite{Mbalsone}), global entanglement~\cite{Dameyer} and some other measures~\cite{Aosterloh0} 
have been used to describe many-body phenomena~\cite{Tcwei}.
However, in general, they are hard to compute and it is therefore not possible to use them for states of arbitrary many-body systems.
Recently, a genuine multiparticle entanglement measure 
called generalized geometric measure (GGM)~\cite{Asende} has been introduced which can be easily 
computed for pure states in arbitrary dimensions and of an arbitrary number of particles. 
It has since been found to be useful to study the genuine multiparty entanglement present in systems like resonating valence
bond states and states of disordered systems~\cite{Rprabhu}. In this paper, we apply the GGM to study quantum spin models, including 
frustrated ones~\cite{Gmisguich,Mrasolt}. Frustration appears in a many-body system if it is not possible to
simultaneously and independently minimize all the interaction terms of the corresponding Hamiltonian~\cite{Mrasolt}. 
The characterization of such systems is typically hard to achieve~\cite{Mrasolt}. We consider the following four classes of quantum 
spin systems -- 
\begin{enumerate}
	\item the quantum one-dimensional (1D) transverse $XY$ model~\cite{Elieb},
	\item the 1D antiferromagnetic $J_1 - J_2$ Heisenberg model~\cite{Ckmajumdar,Srwhite,Rchhajlany,Hjmikeska},
	\item the antiferromagnetic $J_1 - J_2$ model on a two-dimensional (2D) lattice~\cite{Gmisguich}, and
	\item the Shastry-Sutherland model~\cite{Bsshastry}.
\end{enumerate}
The choice of the above models is due to their immense importance in understanding the different exotic 
phases in many-body systems including high-T$_c$ superconductivity~\cite{Pwanderson}. Moreover, the recent 
experimental realizations of such spin models in the laboratories~\cite{Mlewenstein,Kkim,Jstruck,Jzhang}, 
for example, in optical lattice~\cite{Mlewenstein,Ibloch}, trapped ions~\cite{Dleibfried}, photons
~\cite{Xsma}, and nuclear magnetic resonance~\cite{Xpeng}, have led to the interesting possibility of the observation 
of the many-body effects described here in the laboratories. 
Towards the unfolding of such many-body effects, we apply the
multiparty entanglement measure, the GGM, to study the phase diagrams in these models, from a multipartite entanglement perspective.
It is observed that the phase diagrams so obtained, indicate transitions from gapless to gapped phases and vice-versa
in these models.

The approach chosen in the paper is as follows. Entanglement properties have been suspected to be related to a large variety of 
cooperative phenomena in many-body physics. However, the analysis of this suspicion is made difficult by the intractability of most
entanglement measures, especially in the multiparty domain. The generalized geometric measure (GGM) is a recently proposed genuine 
multiparty entanglement measure and is, to our knowledge, the only measure that can be computed for any pure quantum state of an 
arbitrary number of parties in any dimension. We wish to use this fact to our advantage towards analyzing multiparty entanglement 
in many-body systems. The proposal is to use the GGM as an order parameter to detect quantum phase 
transitions. We first use the GGM to check whether it can effectively capture the well-known quantum phase transition in the transverse
XY model. It is to be noted that the quantum transverse XY model can be solved exactly, and it will therefore be satisfying to find that the 
GGM detects the quantum phase transition in this model precisely, without any concern for finite-size effects. Having obtained this 
result, we then look for the possibility of the GGM acting as a detector of quantum phase transitions in the other
models, which are not exactly solvable, and moreover in which the quantum phase transitions are not precisely known by considering the
other order parameters used in the literature.
For comparison, we evaluate the bipartite quantum correlation measures towards detecting the quantum critical points of all the 
above models. We find that the GGM is a better indicator of quantum phase transitions than the bipartite measures, especially 
for  two-dimensional lattice models.

The paper is organized as follows: In Sec. \ref{secggm}, we present a formal definition of the genuine multiparty 
entanglement measure, viz., the GGM, and we show that the same can be expressed in terms of easily computable Schmidt coefficients.
 We also discuss the reasons for choosing GGM over bipartite measures like concurrence and logarithmic negativity
for the present investigations.
Further results are presented in Sec. \ref{secsipnmodels}, where in each subsection, we consider one of the 
quantum spin models. Sections~\ref{anisoXY}, \ref{1dJ1J2}, \ref{2dJ1J2} and \ref{ss} deal, respectively, with the anisotropic $XY$, 
the 1D frustrated $J_1-J_2$ model, the 2D frustrated $J_1-J_2$ model and the Shastry-Sutherland model. In Section~\ref{compare},
we compare the GGM with other bipartite quantum measures of shared quantum systems in terms of detecting QPTs.
Finally, we draw our conclusions in Sec. \ref{secconclusion}.

\section{Generalized Geometric Measure}
\label{secggm}
\noindent In this section, we present a brief description of the GGM, and show that it is efficiently computable for pure quantum 
states of an arbitrary number of parties. 
A pure quantum state  
\(|\psi\rangle_{A_1 A_2 \ldots A_N}\), 
shared between $N$ parties, $A_1, A_2, \ldots, A_N $, is said to be genuinely \(N\)-party entangled, 
if it is not a product across any bipartite partition. 
The GGM of  a pure quantum state  \(|\psi\rangle_{A_1 A_2 \ldots A_N}\) is defined as
\begin{equation}
{\cal E} (  |\psi\rangle_{A_1 A_2 \ldots A_N}) = 1 - \Lambda^2_{\max} ( |\psi\rangle_{A_1 A_2 \ldots A_N}).
\label{GGMdefi}
\end{equation}
Here \(\Lambda_{\max} ( |\psi\rangle_{A_1 A_2 \ldots A_N}) =
\max | \langle \phi|\psi\rangle_{A_1 A_2 \ldots A_N}|\) where the maximization is taken over all pure states \(|\phi\rangle\)
which  are not genuinely \(N\)-party entangled. 

Let us enumerate some of the properties of GGM. 
\begin{enumerate}
\item \({\cal E}\)  is non-vanishing for all  genuine multiparty entangled states, and vanishing for others.
\item  \({\cal E}\)  is monotonically non-increasing under  local (quantum) operations and classical communication.
(The proof follows from the theorem stated below and Ref.~\cite{Manielsen}.)
\end{enumerate}
We now prove a theorem, where we show how the GGM can be expressed in terms of Schmidt coefficients.\\
 
\noindent\textbf{Theorem:} The generalized geometric measure (GGM) can be expressed as 
 \begin{equation}
{\cal E}(|\psi\rangle) = 1 - \max \{\lambda^2_{{\cal A}: {\cal B}} | {\cal A} \cup {\cal B} = \{A_1,\ldots,A_N\}, {\cal A} \cap  {\cal B} = \emptyset\},
\label{GGMsimpl}
\end{equation}
where \(\lambda_{{\cal A}: {\cal B}}\) is the maximal Schmidt coefficient of  \(|\psi\rangle_{A_1 A_2 \ldots A_N}\) in the \({\cal A}: {\cal B}\) bipartite split.\\

 \noindent \texttt{Proof:}
 The maximization involved in the definition of GGM, given in Eq.~(\ref{GGMdefi}) is over 
 all N-party pure quantum states \(|\phi\rangle_{A_1 A_2 \ldots A_N}\)
 that are not genuinely multiparty entangled.
 The square of \(\Lambda_{\max}(|\psi\rangle_{A_1 A_2 \ldots A_N})\) can 
 be interpreted as the Born probability of an outcome in some quantum measurement on the multiparty quantum state \(|\psi\rangle_{A_1 A_2 \ldots A_N}\). 
 Since entangled quantum measurements are, in general, better than the product ones for any set of the subsystems involved, the maximization needs to be carried out only in 
 a partition of \(A_1, A_2, \ldots, A_N\) into two parts.
 In other words, the maximization in \(\max |\langle \phi | \psi_{A_1 A_2 \ldots A_N} \rangle|\) is performed over the \(|\phi\rangle_{A_1 A_2 \ldots A_N}\) that are
  product across some bipartition, say, \({\cal A}: {\cal B}\). This is exactly the  maximal Schmidt coefficient, \(\lambda_{{\cal A}: {\cal B}}\), 
 of the state \(|\psi\rangle_{A_1 A_2 \ldots A_N}\) in the \({\cal A}: {\cal B}\) bipartite split.
 Note that \(\lambda_{{\cal A}: {\cal B}}\) are increasing under LOCC~\cite{Manielsen}. This immediately implies that GGM ${(\cal E)}$ is non-increasing under LOCC.
And, \(\Lambda_{\max}(|\psi\rangle_{A_1 A_2 \ldots A_N})\) is  the maximum of all such maximal Schmidt coefficients in all possible bipartite splits of the N parties.
Hence, the theorem.~\hfill $\blacksquare$\\

 The theorem makes it possible to calculate the GGM for any pure state of an arbitrary number of parties in arbitrary dimensions. 
 This is due to the fact that the definition of GGM, given in Eq.~(\ref{GGMdefi}), reduces to the calculation of 
 squares of the maximal Schmidt coefficients across all bipartitions, as given in Eq.~(\ref{GGMsimpl}). So, for example, for 
 calculating the GGM of a four-party symmetric state $|\psi\rangle_{ABCD}$, we have to consider $|\psi\rangle$ in
 the A:BCD and AB:CD partitions, and find the maximal Schmidt coefficients in these partitions. The GGM of $|\psi\rangle_{ABCD}$
 is then 1-$\lambda^2$, where $\lambda$ is the highest of these maximal Schmidt coefficients.
 
 There are a large number of concepts that can be used to quantify entanglement. The reasons that we use the GGM here
 are as follows. It is widely believed that entanglement of many-body systems has  an important bearing on the 
 cooperative physical phenomena in those systems. Since a large number of particles (sub-systems) are necessary for generating 
 such effects, it is plausible that it is the multiparty entanglement that would better reveal the positions and characteristics
 of these cooperative phenomena. This belief is reinforced by the recent results demonstrating that bipartite entanglement 
 measures like concurrence and logarithmic negativity cannot reliably capture the position of quantum phase transitions in some
 systems~\cite{Mfyang,Xfqian}. 
 It is therefore all the more natural to look out for multipartite entanglement
 measures to investigate such cooperative phenomena in many-body systems.
The generalized geometric measure is, to our knowledge, the only measure of genuine multiparty entanglement that can be computed 
for any pure quantum state of an arbitrary number of parties in any dimensions. This led us to use it to study transitions in important 
many-body systems.

The scaling of the von Neumann entropy is a way to understand the multiparty entanglement properties 
of quantum many-body states. In this case, the corresponding state $|\psi_N\rangle$ of, say, $N$ spin-$\frac{1}{2}$ particles, is first divided 
into two parts, one of which consists of $L$ particles while the other consists of the rest. Then the von Neumann entropy, $S_L$, of the
subsystem of $L$ particles, corresponds to the entanglement of $|\psi_N\rangle$ in the $L:N-L$ bipartition. 
Note here that the entropy has to be found by using all the Schmidt 
coefficients of the density matrix of the subsystem.

In a path-breaking series of papers, a change in the scaling law, {i.e.}, the behavior of $S_L$ versus $L$, has been proposed to be an 
order parameter to detect quantum phase transitions~\cite{Lamico, qwerty}. The scaling of local von Neumann entropy however has important drawbacks. 
Firstly, local von Neumann entropy is no more a measure of 
entanglement for mixed quantum shared states. Second and perhaps more important, is the fact that  while scanning over a system parameter 
to ``pin down'' a phase transition, it is difficult, practically, to detect a change in the scaling law, especially for systems where 
finite-size calculations are essential, due to analytical intractability. This is because at every point in parameter space, we obtain a 
function. Scanning over the parameter space, we get a family of functions. Detecting phase transitions by looking for changes in the 
functional form is a mathematically difficult problem, especially if the functional forms are all approximate to begin with.

Both these drawbacks are overcome by considering the GGM. The GGM is well-defined for both pure and mixed states. Moreover, scanning over the 
parameter space, we obtain a surface defined on the parameter space, because for every point of the parameter space, the GGM at that point is
a real number. We then identify phase transitions with some drastic change in behavior of this surface.

Furthermore, 
to calculate the GGM, one should consider all 
possible bipartitions of the $N$-partite state (in the maximization given in Eq.~(\ref{GGMsimpl})) which include bipartitions whose parts are not separately 
connected. On the other hand, 
in considerations of the scaling of von Neumann entropy, one usually considers connected clusters of lattice sites. However, there are important exceptions~\cite{arealaw} where researchers 
have gone beyond this usual practice.

\section{BEHAVIOR OF GGM NEAR GAPLESS - GAPPED QUANTUM TRANSITIONS}
\label{secsipnmodels}

\noindent In this section, we consider a series of paradigmatic quantum spin systems.
 They are taken up one by one in the different subsections. Each subsection begins with a brief description
 of the Hamiltonian corresponding to the quantum system under study.
 Subsequently, we study the behavior of the ground state
of these models, and investigate the advantage of considering the genuine 
multipartite entanglement measure in these models.

\subsection{Anisotropic  XY model}
\label{anisoXY}
\noindent The Hamiltonian for the one-dimensional anisotropic quantum $XY$ model of $N$ quantum spin-half particles, arranged in an 1D array, 
is given by~\cite{Elieb}
\begin{equation}
\label{eq_XY_H}
 H_{XY} = \frac{J}{2} \left(\sum_{i=1}^{N} (1 + \gamma) \sigma^x_i \sigma^x_{i+1} + 
 (1 - \gamma) \sigma^y_i \sigma^y_{i+1}\right) + h \sum_{i=1}^{N} \sigma_i^z,
\end{equation}
where $J$, which has the units of energy, is of the same order as the coupling constant for the nearest neighbor interaction, $\gamma \in (0,1]$ is the 
(dimensionless) anisotropy 
parameter, \(\sigma\)'s are the Pauli spin matrices, 
and \(h\), which again has the units of energy, represents the external transverse magnetic field applied across the system. In all the models 
considered in this paper, we impose the periodic boundary condition. 
The quantum XY Hamiltonian can be diagonalized by applying Jordan-Wigner, Fourier, and Bogoliubov transformations~\cite{Elieb}. 
At zero temperature, it undergoes 
a quantum phase transition driven by the transverse magnetic field at $\lambda \equiv h/J=1$. Moreover, it is also 
known that the model
is gapped for all field strengths except at the point where the quantum phase transition occurs. Such 
transitions have been detected by using bipartite entanglement measures like concurrence and multipartite entanglement 
measures like geometric measure~\cite{Tcwei,Aosterloh}. We investigate the behavior of the genuine multipartite
entanglement measure viz., the GGM, of the ground state, when it crosses
from one gapped phase to another, through the gapless
point. 

\begin{figure}
\centerline{
\includegraphics[height=8.5cm,width=6.5cm,angle=-90]{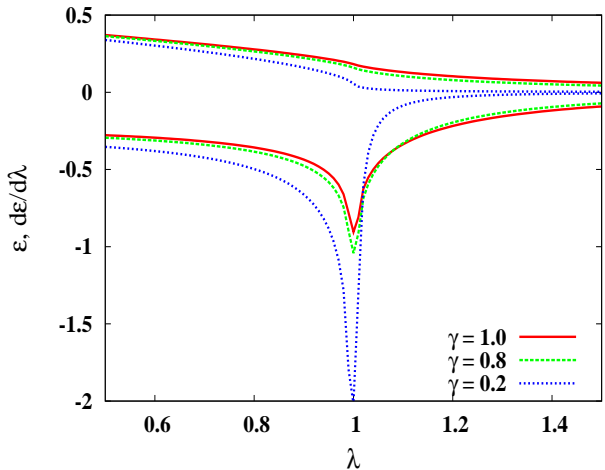}}
\caption{ (Color online) GGM of the transverse XY model. The GGM and its derivative (both dimensionless)
are plotted on the vertical axis for the anisotropic transverse XY model for different anisotropy parameters
\(\gamma\), against the dimensionless system parameter \(\lambda\) on the horizontal axis. The plots are for
the Ising (\(\gamma = 1\)), \(\gamma = 0.8\), and \(\gamma =  0.2\) models. The derivatives of the GGM diverge 
at the quantum critical point \(\lambda  = 1\). The cluster of three upper curves are for the GGM, while the
lower ones are for their derivatives. For the purpose of the figure, we have used the eigenvalues corresponding 
to the single, two-, and three-site density matrices of the ground state.
}
\label{XY-GGM}
\end{figure}

The ground state of the system represented by the quantum XY Hamiltonian, as given in Eq. (\ref{eq_XY_H}), can be
analytically obtained by using Majorana fermions and it is also possible to get the eigenvalues of the 
local density matrices corresponding to the ground state, in different bipartitions~\cite{Elieb,Lamico}.
The local density matrix corresponding to \(L\) consecutive sites can be obtained by calculating their correlators and magnetizations. 
The largest eigenvalue of the local density matrix corresponding 
to $L$ sites, where $1\leq L \leq N/2$, when subtracted from 1, gives the GGM. 
We have assumed here that the density matrices corresponding to non-consecutive sites
do not produce significant eigenvalues to contribute to the GGM.
We have checked that this assumption remains valid for moderate-sized finite XY chains. 
The assumption is intuitively satisfactory as we are dealing with a nearest neighbor interaction
model.
In Fig.~\ref{XY-GGM}, we have plotted the GGM and the derivative of the GGM for the ground state of the $XY$ model 
with respect to the driving parameter, $\lambda$, for different values of the anisotropic 
constant $\gamma$. 
The divergence of the derivative of 
GGM 
captures the presence of the quantum phase transition at $\lambda=1$.
When $\gamma=1$, which corresponds to the Ising model, the genuine multipartite entanglement is maximum when 
 compared to the systems with lower values of $\gamma$.

\subsection{1D Frustrated \(J_1-J_2\) Model}
\label{1dJ1J2}
\noindent We will now consider the frustrated  one-dimensional \(J_1-J_2\) Heisenberg model,
in which the nearest neighbor couplings, \(J_1\), and 
the next-nearest neighbor couplings, \(J_2\), are both antiferromagnetic. 
It was found that solid state systems like \(\mbox{SrCuO}_{2}\) can be described 
by this model~\cite{Mmatsuda}. Moreover, advances in the field of cold atomic systems promise to 
create and control such models in the laboratory~\cite{Mlewenstein}.
The Hamiltonian of this model, with \(N\) lattice sites on a chain, can be written as
\begin{equation}
 H_{1D} = J_1 \sum_{i=1}^{N} \vec{\sigma}_i \cdot \vec{\sigma}_{i+1}
 + J_2 \sum_{i=1}^N \vec{\sigma}_i \cdot \vec{\sigma}_{i+2},
\end{equation}
where \(J_1\) and \(J_2\) are both positive. 
In the parameter space,  \(\alpha \equiv J_2/J_1 = 0.5\) is known as the Majumdar-Ghosh point, and the system is 
highly frustrated there.  
For an even number of sites, the ground state at 
the Majumdar-Ghosh point 
is doubly degenerate, and the ground state manifold 
is spanned by the two dimers
\(|\psi_{MG}^{\pm}\rangle = \Pi_{i=1}^{N/2} (|0\rangle_{2i} |1\rangle_{2i \pm 1} -  |1\rangle_{2i} |0\rangle_{2i \pm 1})\).
Note that the 
model is gapped at this point~\cite{Ckmajumdar}. 
For \(\alpha=0\), the Hamiltonian reduces to the spin\(-\frac{1}{2}\) Heisenberg antiferromagnet and hence the ground state, which is a spin fluid state having  
gapless excitations~\cite{Rbgriffiths}, can be studied by Bethe ansatz~\cite{Hbethe}. At other points, the ground state and the energy gap of this model
were considered by using exact diagonalization, density matrix renormalization group method, bosonization technique, etc~\cite{Hjmikeska}.
It is known that  at \(\alpha \approx 0.2411\), a phase transition from fluid 
to  dimerization occurs~\cite{Fdmhaldane}. In the weakly frustrated regime, $0 < \alpha \lesssim 0.24$, the system is
gapless, and therefore critical~\cite{Ckmajumdar,Srwhite}. The system enters a dimerized regime, and is gapped, for higher values
of the coupling parameter. 

\begin{figure}[hbpt]
\vspace{-10pt}
\centerline{
\hspace{-3.3mm}
\rotatebox{-90}{\epsfxsize=6cm\epsfbox{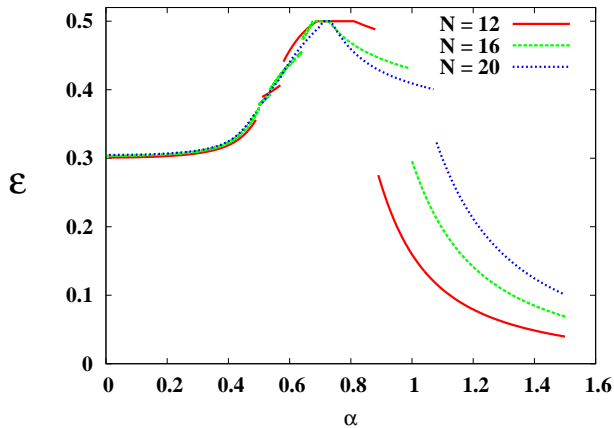}}}
\caption{
(Color online.) GGM for the 1D frustrated \(J_1-J_2\) Model. The GGM (dimensionless) is plotted on the vertical axis 
and system parameter $\alpha$ (dimensionless) is plotted along horizontal axis. 
Note that the GGM starts to increase from its almost constant value of $\sim 0.3$ in the gapless region, 
around $\alpha \approx 0.3$.
We will take a closer look at the figure in Sec.~\ref{compare}.
}
 \label{ggm-bggm-1d}
\end{figure}

 We perform exact diagonalization of the Hamiltonian using TITPACK ver.~2 developed by H. Nishimori~\cite{Hnishimori} 
to find the ground state and then compute the GGM.
In Fig.~\ref{ggm-bggm-1d}, the GGM is almost constant with respect to the driving parameter in the region when the system is gapless. 
It begins to increase with respect to the driving parameter near the phase transition point. 
However, due to the small system size that is accesible to study, 
it is difficult to locate the exact QPT point from the figure.
Our investigation nevertheless reveals the behavior of multisite entanglement in the relevant parameter space
of the finite-size 1D frustrated \(J_1-J_2\) model.

The discontinuities in the GGM 
curves
are arguably due to the finite and small system sizes. 
Note that the amounts of the discontinuities in Fig.~\ref{ggm-bggm-1d} decrease with increase in the system size from $N=12$ to $N=20$
and it is plausible that they will disappear for larger systems.
These discontinuities appear at avoided level crossings of the two lowest eigenvalues of the system Hamiltonian.
Note that the behavior of entanglement entropy is also similar for this model (see Fig.~7 of Ref.~\cite{Rchhajlany}).
Also, note that the 
GGM curves asymptotically go to zero for very high values of the driving parameter $\alpha$, as then the spin chain decouples 
into two spin chains with nearest neighbor interaction couplings, $J_2$. 

\subsection{2D Frustrated \(J_1-J_2\) Model}
\label{2dJ1J2}

\noindent We now consider an arrangement of quantum spin$-\frac{1}{2}$ particles on a 2D square lattice, where nearest neighbor spins (both vertical and horizontal)
 on the lattice are coupled by Heisenberg interactions, with coupling strengths \(J_1\), and where 
all diagonal spins are coupled by the same interactions, with coupling strengths \(J_2\). Both $J_1$ and $J_2$ are considered to be positive. 
The model has attracted a lot of attention~\cite{Pchandra} 
due to its connection 
with high \(T_c\)-superconductors and 
its similarity with magnetic materials 
like \(\mbox{Li}_2 \mbox{VOSiO}_4\) and  \(\mbox{Li}_{2}\mbox{VOGeO}_4\)~\cite{Rmelzi}. 
Although the different phases of the ground state of this model have been predicted by different
 numerical as well as approximate analytical methods~\cite{Jrichter}, some debates remain.
The Hamiltonian of the system is given by 
\begin{equation}
 H_{2D} = J_1 \sum_{\langle NN \rangle} \vec{\sigma}_i \cdot \vec{\sigma}_{j} + 
J_2 \sum_{\langle \scriptsize{\mbox{diagonals}} \rangle} \vec{\sigma}_i \cdot \vec{\sigma}_{j},
\end{equation}
where \(J_1\) and \(J_2\) are antiferromagnetic.

\begin{figure}[]
\vspace{-10pt}
\centerline{
\hspace{-3.3mm}
\rotatebox{-90}{\epsfxsize=6cm\epsfbox{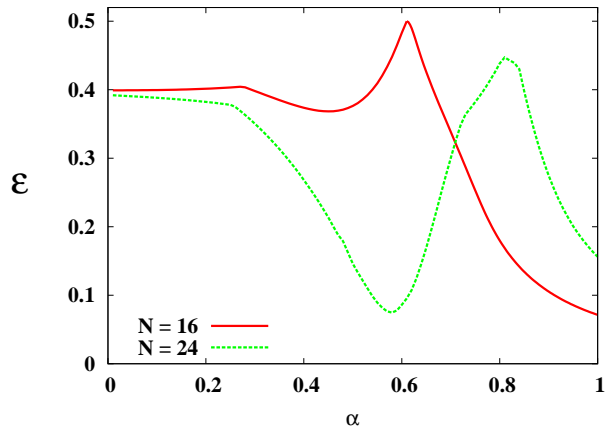}}}
\caption{
(Color online.) GGM for the 2D frustrated \(J_1-J_2\) Model. The GGM (vertical) is plotted against the system parameter $\alpha$ (horizontal). 
Both the quantities are dimensionless. 
}
\label{ggm-2d}
\end{figure}

In the classical limit, only a first-order phase transition from N{\' e}el 
to collinear at \(\alpha \equiv J_2/J_1 = 0.5\) is exhibited by this model.
The nature of the phase diagram changes when  quantum fluctuations  are present, and in this case, 
the exact phase boundaries are not known. Based on exact diagonalization, series expansion 
methods, field-theory methods~\cite{Jrichter}, etc., one expects that there are two long range ordered (LRO) ground state
phases in the system, that are separated by quantum paramagnetic phases without LRO.
These investigations predict that the first transition from
N{\' e}el to dimer occurs at \(\alpha=\alpha^c_1 \in (0.3,0.45)\) while 
dimer to collinear transition happens at \(\alpha=\alpha^c_2 \in (0.6,0.7)\). 
There are 
proposals of detecting these phases in the laboratory, and they
demand a precise quantification of the low temperature phase diagram of this model.

We investigate the behavior of genuine multipartite entanglement of the ground state 
and study its effectiveness to detect the transitions present in the system. To obtain the ground states, 
we use exact diagonalization technique as mentioned in the preceding subsection.
In Fig.~\ref{ggm-2d}, we plot the GGM as a function of the driving parameter $\alpha$. 
The non-analyticity of the GGM with respect to the driving parameter $\alpha$ indicates a N{\' e}el (gapless) to dimer (gapped) 
transition occurs at $\alpha \approx 0.27$ for $N=16\) and 
$\alpha \approx 0.25$ for $N=24$, while 
the dimer to collinear transition point is in the range $\alpha \in (0.61,0.62)$
 for $N=16$
and $\alpha \approx 0.81$ for $N=24$. 
In case of the second transition, the \(N=24\) case predicts a transition at a point that is somewhat 
away from previous predictions. We believe that this is due to the fact that \(24\) is not a perfect square. 
The results indicate that the second transition is more sensitive to the lattice structure, for such small systems.  
Just like in the 1D case, the GGM curves asymptotically go to zero for high $\alpha$. This is because 
the spin lattice decouples into two spin lattices with nearest neighbor 
 interaction couplings, $J_2$, for very high values of the driving parameter $\alpha$.

\subsection{The Shastry-Sutherland Model}
\label{ss}

\begin{figure}[h]
\centerline{
\epsfig{figure=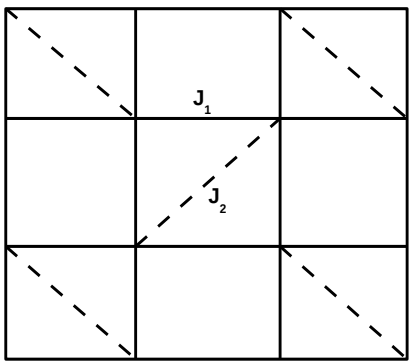, height=.2\textheight,width=0.3\textwidth}}
\caption{The Shastry-Sutherland lattice. The solid lines represent the nearest neighbor interactions with coupling strength $J_1$ 
between the lattice sites and the ones joined by the dashed lines represent next neighbor interactions with coupling strength $J_2$.}
\label{ss-model}
\end{figure}

\noindent In this section, we study the entanglement properties of systems where the interaction between particles 
can be modeled by the Shastry-Sutherland Hamiltonian~\cite{Bsshastry}. In the insulators like SrCu$_2$(BO$_3$)$_2$, 
the low-energy spin excitations reside on the spin-half copper ions which lie in two dimensional layers decoupled from each other.
The antiferromagnetic exchange 
couplings between the Cu ions is identical to the Shastry-Sutherland Hamiltonian. The lattice with schematic interactions, for this model, is given in 
Fig.~\ref{ss-model} and the Hamiltonian is given by
\begin{equation}
\label{ss-hamiltonian}
 H_{SS}=J_1 \displaystyle \sum_{NN} \vec{\sigma}_i.\vec{\sigma}_j + J_2 \displaystyle \sum_{NNN} \vec{\sigma}_i.\vec{\sigma}_j.
\end{equation}
Here, $J_1 (>0)$ corresponds to nearest neighbor interaction (indicated by solid lines in Fig. \ref{ss-model}) and $J_2 (>0)$ corresponds to specific next nearest neighbors (indicated by broken lines in Fig.~\ref{ss-model}).
It is known that a simple product of singlet pairs, on the diagonal links, 
is the ground state of $H_{SS}$ for sufficiently large $\alpha \equiv J_2/J_1$. It has been previously reported~\cite{Chchung,Malbrecht} 
that the system undergoes two quantum 
phase transitions driven by the quantum fluctuations: one is from N{\' e}el to an intermediate phase 
and the other one is from that intermediate phase to dimer. The nature
of the intermediate phase is not yet clearly understood~\cite{Wzheng}.

\begin{figure}[h]
\vspace*{.5cm}
\centerline{
\hspace{-3.3mm}
\rotatebox{-90}{\epsfxsize=6cm\epsfbox{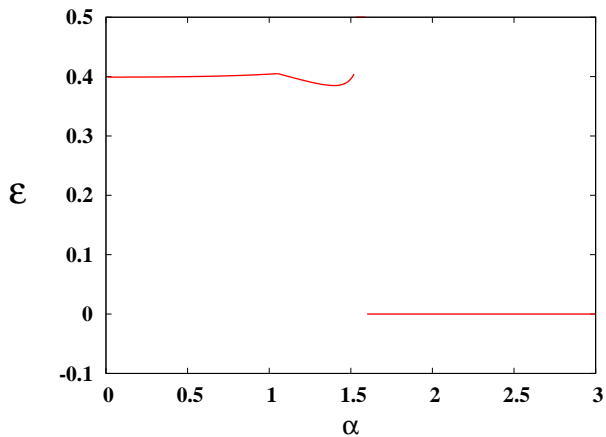}}}
\caption{(Color online.) GGM for the Shastry-Sutherland antiferromagnet. GGM (on vertical axis) is plotted with respect to the 
system parameter $\alpha$ (on horizontal axis) for 16 spins. 
Both the quantities are dimensionless.}
\label{ss-ggm-16}
\end{figure}

In Fig.~\ref{ss-ggm-16}, we plot the GGM as a function of  
$\alpha$ for 16 particles on a square lattice interacting via the Shastry-Sutherland Hamiltonian. 
There are clear signatures of these phase transitions in the figure at $\alpha \approx 1.05$ and 
for $\alpha \approx 1.53$, as have been predicted by other methods \cite{Chchung,Malbrecht}. 
For very high $J_2$, the system consists of isolated dimers and the multisite entanglement vanishes. 
In Fig.~\ref{ss-ggm-16}, we find that the GGM becomes zero at $\alpha \ge 1.53$. 
The GGM curve is non-analytic at the phase transition points in this model as well.
Due to the computational constraints we are
able to report our finding only for $N = 16$. 
More specifically, we expect that 
studying the Shastry-Sutherland model requires an exact square structure with even numbers of spins on each side, and hence the next relevant lattice size is  \(N=36\). 

\section{Comparison of GGM with bipartite quantum measures}
\label{compare}
\noindent Let us now compare the efficiency to detect the critical points
by the genuine multipartite entanglement measure, GGM, with those by 
bipartite quantum characteristics in shared states of quantum spin models. We  
focus on the $N=16$ case, in all the non-integrable models considered in this paper. We calculate
the concurrence~\cite{Wwooters}, logarithmic negativity~\cite{VidalWerner}, quantum discord~\cite{Lhenderson},
and the shared purity~\cite{Abiswas} for the nearest neighbor density matrices in these models. 

\begin{figure}[hbpt]
\begin{center}
\includegraphics[keepaspectratio=false, width=6cm, height=8cm, angle=-90]{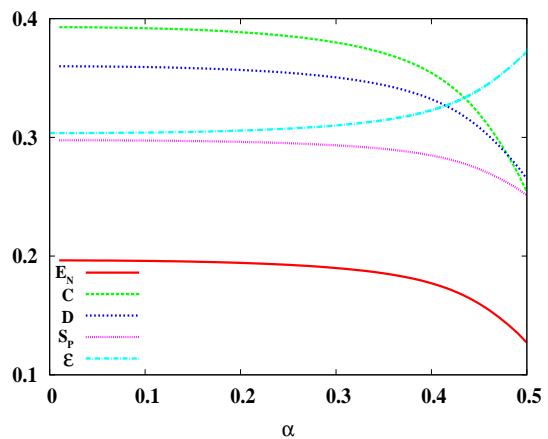}
\caption{(Color online.) Logarithmic negativity, concurrence, quantum discord, shared purity, and GGM,
 with respect to the driving parameter $\alpha$, for the 1D $J_1-J_2$ Hamiltonian consisting of 16 spins.
 Note that all quantities begin to deviate from their $\alpha=0$ values above $\alpha\gtrsim0.25$.
 The horizontal axis is dimensionless. For the vertical axis, logarithmic negativity and concurrence 
 are measured in ebits, quantum discord in bits, while shared purity and GGM are dimensionless.
 We have denoted the quantum discord as $D$ here, underlining the symmetric nature of the two-spin state.}
\label{fig:1dcmp}
 \end{center}
\end{figure}

The bipartite quantum correlation measures can be broadly classified into two classes --
(i) the entanglement-separability paradigm and (ii) the information-theoretic paradigm.
The concurrence and logarithmic negativity
are measures of bipartite entanglement belonging to the entanglement-separability paradigm of quantum correlation measures, 
 while quantum discord, which quantifies bipartite quantum correlation, belongs to the information-theoretic paradigm. The 
shared purity is a newly defined quantum characteristic of shared multipartite quantum systems which is different from 
quantum correlations. We briefly discuss the measures below.
\begin{center}
 {\bf Concurrence}
\end{center}
For any two-qubit state, \(\rho_{AB}\), the concurrence~\cite{Wwooters} is given by  
${\cal C}(\rho_{AB})=\mbox{max}\{0,\lambda_1-\lambda_2-\lambda_3-\lambda_4\}$, where the
$\lambda_i$'s are the square roots of the eigenvalues of $\rho_{AB}\tilde{\rho}_{AB}$ in decreasing order and 
$\tilde{\rho}_{AB}=(\sigma_y\otimes\sigma_y)\rho_{AB}^*(\sigma_y\otimes\sigma_y)$, with $\sigma_y$ being the Pauli spin matrix.
\begin{center}
 {\bf Logarithmic negativity}
\end{center}
The negativity of a bipartite quantum state $\rho_{AB}$, denoted by $\cal N(\rho_{AB})$ is defined as the sum of the 
negative eigenvalues of $\rho_{AB}^{T_A(T_B)}$, where $\rho_{AB}^{T_A(T_B)}$ denotes the partial transpose of $\rho_{AB}$
with respect to $A(B)$. Then the logarithmic negativity~\cite{VidalWerner} of $\rho_{AB}$ is defined as
\begin{equation}
E_{\cal N}(\rho_{AB}) = \log_2 [2 {\cal N}(\rho_{AB}) + 1].
\label{eq:LN}
\end{equation}
The positivity of logarithmic negativity guarantees that the state is entangled.
\begin{center}
 {\bf Quantum Discord}
\end{center}
Quantum discord~\cite{Lhenderson} for a bipartite state $\rho_{AB}$ is defined as the difference between the total correlation and the classical correlation of the state. 
The total correlation, defined as the quantum mutual information of  \(\rho_{AB}\), is given by
\begin{equation}
\label{qmi}
\mathcal{I}(\rho_{AB})= S(\rho_A)+ S(\rho_B)- S(\rho_{AB}),
\end{equation}
where $S(\sigma)= - \mbox{tr} (\sigma \log_2 \sigma)$ is the von Neumann entropy of the quantum state \(\sigma\). The classical correlation, based on the conditional entropy,
is defined as
\begin{equation}
\label{eq:classical}
 \overleftarrow{\cal J}(\rho_{AB}) = S(\rho_A) - S(\rho_{A|B}). 
\end{equation}
Here,
\begin{equation}
 S(\rho_{A|B}) = \min_{\{B_i\}} \sum_i p_i S(\rho_{A|i})
\end{equation}
is the conditional entropy of \(\rho_{AB}\), conditioned on a measurement performed by \(B\) with a rank-one
projection-valued operators \(\{B_i\}\),
producing the states  
\(\rho_{A|i} = \frac{1}{p_i} \mbox{tr}_B[(\mathbb{I}_A \otimes B_i) \rho (\mathbb{I}_A \otimes B_i)]\), 
with probability \(p_i = \mbox{tr}_{AB}[(\mathbb{I}_A \otimes B_i) \rho (\mathbb{I}_A \otimes B_i)]\).
\(\mathbb{I}\) is the identity operator on the Hilbert space of \(A\). Hence the discord can be calculated as
\cite{Lhenderson}
\begin{equation}
\label{eq:discord}
\overleftarrow{\cal D}(\rho_{AB})= {\cal I}(\rho_{AB}) - \overleftarrow{\cal J}(\rho_{AB}).
\end{equation}
Here, the superscript ``$\overleftarrow{}$" on ${\cal J}(\rho_{AB})$ and ${\cal D}(\rho_{AB})$ indicates that
the measurement is performed on the subsystem $B$ of the state $\rho_{AB}$. Similarly, if measurement is performed on the subsystem $A$ of the state $\rho_{AB}$, one can define a quantum discord as
 \begin{equation}
\label{eq:discordA}
\overrightarrow{\cal D}(\rho_{AB})= {\cal I}(\rho_{AB}) - \overrightarrow{\cal J}(\rho_{AB}).
\end{equation}
In our case, $\overleftarrow{\cal D}(\rho_{AB})=\overrightarrow{\cal D}(\rho_{AB})$, which is a 
consequence of the periodic boundary condition used for our analysis.
\begin{center}
 {\bf Shared Purity}
\end{center}
Shared purity~\cite{Abiswas} is the difference between the ``global'' and ``local'' fidelities of an arbitrary (pure or mixed) state $\rho$.
The global fidelity is a measure of the minimum distance of 
the state $\rho$ from a globally pure state while the local fidelity is a measure of the minimum distance of $\rho$ from a 
locally pure state. The ``global fidelity'' of an $N$-party arbitrary
(pure or mixed) quantum state, $\rho_{1\ldots N}$, on $\cal{H}=\mathbb{C}^{d_1}\otimes\ldots\otimes \mathbb{C}^{d_N}$, is defined as
\begin{equation} 
 F_G=\displaystyle \max_{\lbrace|\phi\rangle_{1\ldots N}\in \mathcal{H}\rbrace} {_{1\ldots N}\langle\phi|\rho_{1\ldots N}|\phi\rangle_{1\ldots N}},
\end{equation}
where the maximization is performed over all elements (pure states) of $\mathcal{H}$. And the ``local fidelity'', of the same state 
is defined as
\begin{equation} 
 F_L=\displaystyle \max_{\lbrace|\phi\rangle_{1\ldots N} \in S\rbrace} 
{_{1\ldots N}\langle\phi|\rho_{{1}\ldots{N}}|\phi\rangle_{1\ldots N}},
\end{equation}
where the maximization is carried out over a certain set $S$, of pure product states.
For bipartite systems, the set $S$ consists of all pure product states.
The shared purity denoted by $S_P$ is defined as 
\begin{equation}
 S_P=F_G-F_L.
\end{equation}
\begin{center}
 {\bf Comparison}
\end{center}
\noindent In Fig.~\ref{fig:1dcmp}, we compare the concurrence, logarithmic negativity, quantum discord, and 
shared purity  with the GGM, calculated for the ground state of the 1D $J_1-J_2$ Hamiltonian consisting of 16 spins,
with respect to the
driving parameter $\alpha$. The GGM is calculated for the 16-spin ground state while the other quantities are 
calculated for the nearest neighbor two-spin reduced density matrix of the same 16-spin state.
The system remains in the gapless phase for $\alpha\lesssim0.24$. It can be seen from the figure that all the quantities
 begin to deviate from their $\alpha=0$ values when $\alpha\gtrsim0.25$. 
 Note that the $\alpha=0$ case corresponds to the isotropic Heisenberg nearest neighbor antiferromagnetic chain.
 Although there is no definite signature of a QPT from
 any 
 of the quantum measures, the critical point can be estimated to lie at $\alpha\approx0.25$
 by comparing with the $\alpha=0$ case. Note that entanglement 
 entropy was also used 
 to estimate the quantum critical point by exact diagonalization in Ref.~\cite{Rchhajlany}. 
 The quantity plotted there begins to deviate from its value at $\alpha=0$ when $\alpha\gtrsim0.25$.
 Here too, there is no clear signal at the QPT. However, the critical point was estimated to be at $\alpha\approx0.25$.
 In Ref.~\cite{Reryigit}, a multipartite entanglement measure, the global entanglement, was used to study the 1D $J_1-J_2$ Hamiltonian.
 There was no clear signature of the QPT there either.
 It should be added however that the studies, despite not pinning down the QPT in the 1D $J_1-J_2$ model, 
 does serve the important purpose of studying quantum correlation properties around this elusive QPT.
 
\begin{figure}[hbpt]
\begin{center}
\includegraphics[keepaspectratio=false, width=8cm, height=6cm, angle=0]{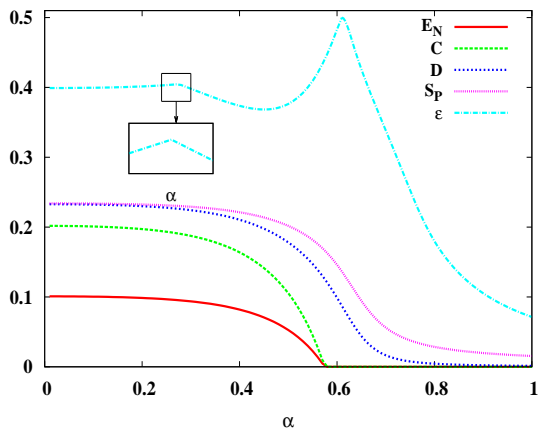}
\caption{(Color online.) Logarithmic negativity, concurrence, quantum discord, shared purity, and GGM,
 with respect to the driving parameter $\alpha$, for the $N=16$ 2D $J_1-J_2$ Hamiltonian.
 A magnified portion of the GGM curve signalling a QPT is plotted in the inset.
 The horizontal axis is dimensionless. All other dimensions and notations are as in Fig.~\ref{fig:1dcmp}.}
\label{fig:2dcmp}
 \end{center}
\end{figure}

 In Fig.~\ref{fig:2dcmp}, we plot the concurrence, logarithmic negativity, quantum discord, and 
shared purity, along with GGM, calculated for the ground state of the 16-spin 2D $J_1-J_2$ Hamiltonian, with respect to the
driving parameter $\alpha$. 
Again we compare the GGM with the bipartite measures.
The GGM clearly signals both the QPTs present in this model by virtue of the discontinuity of the derivative of the GGM
with respect to $\alpha$ at the quantum critical points. The bipartite entanglement measures, viz. concurrence and logarithmic
negativity signal the second QPT at around $\alpha\approx0.58$, where these quantities vanish. Quantum discord and shared 
purity also signal the second critical point at $\alpha\approx0.6$, where the derivatives of these quantities with respect 
to $\alpha$ are minimum. The bipartite measures do not conclusively detect the first quantum critical point. However,
all these bipartite measures begin to deviate from their values at $\alpha=0$, when $\alpha\gtrsim0.3$.
It is clear that the multiparty measure is more efficient in this case than the bipartite measures in
identifying quantum critical points.

\begin{figure}[hbpt]
\begin{center}
\includegraphics[keepaspectratio=false, width=8cm, height=6cm, angle=0]{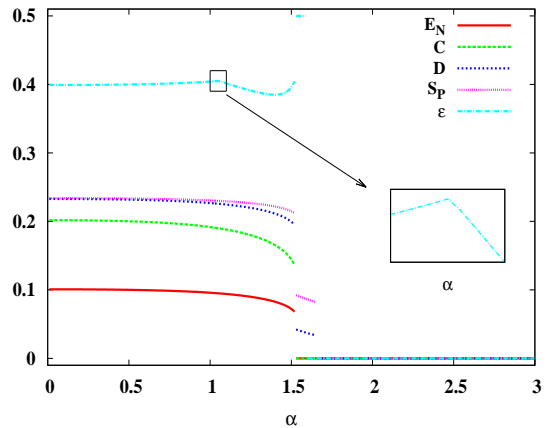}
\caption{(Color online.) Logarithmic negativity, concurrence, quantum discord, shared purity, and GGM,
 with respect to the driving parameter $\alpha$, for the $N=16$ Shastry-Sutherland Hamiltonian.
 A magnified portion of the GGM curve signalling a QPT is plotted in the inset.
 The horizontal axis is dimensionless. All other dimensions and notations are as in Fig.~\ref{fig:1dcmp}.}
\label{fig:sscmp}
 \end{center}
\end{figure}

In Fig.~\ref{fig:sscmp}, we again plot the same 
bipartite measures and compare with the
GGM, calculated for the ground state of the 16-spin Shastry-Sutherland Hamiltonian, with respect to the
driving parameter $\alpha$. 
The derivative of the GGM is discontinuous at both the quantum critical points, while
the bipartite measures can only signal the second quantum critical point at $\alpha\approx1.5$,
 above which they vanish.
There is only a slight indication of the first QPT at $\alpha\approx1$ above which the bipartite 
measures begin to deviate from 
their values at $\alpha=0$. We therefore again find that the multiparty entanglement measure is
a better detector of quantum phase transitions than the bipartite measures.

\section{Conclusions}
\label{secconclusion}

An important classification scheme for multipartite entangled quantum states is according to their separability
in different partitions. The complexity of such a classification makes it difficult to obtain a unique multiparty
entanglement measure, even for pure quantum states. A comparison with the situation for mixed bipartite states
is relevant here. While the entanglement of pure bipartite states can be uniquely characterized by a single 
entanglement measure, a variety of different measures exist for mixed bipartite states. In the case of multiparty
quantum states, one can analogously have ``pure'' and ``non-pure'' forms of entanglement, corresponding 
respectively to genuine multiparty entanglement and ones that are not genuine. In this sense, the generalized 
geometric measure, which is a measure of genuine multiparty entanglement, quantifies the “pure” form of multiparty 
entanglement that is present in a multiparty quantum state. 

We employed this measure to predict phase diagrams in quantum many-body systems. We began by 
using the measure to detect the quantum fluctuation driven phase transition in an exactly solvable model, 
viz. the quantum XY model. We subsequently applied the generalized geometric measure to prototype frustrated 
quantum spin models, in the one-dimensional antiferromagnetic $J_1 - J_2$ model, the two-dimensional antiferromagnetic 
$J_1 - J_2$ model and the Shastry-Sutherland model. The ground states and the corresponding phase diagrams, for the 
frustrated models, are not known exactly, although there have been several predictions by different methods. 
We use exact diagonalization techniques to investigate the multipartite entanglement of the 
ground states of the frustrated models. 
The GGM is non-analytic or its derivative is divergent at the quantum phase transition points in all the models
except the one dimensional $J_1-J_2$ model. For the one dimensional $J_1-J_2$ model, the fluid-dimer transition there
is accompanied by a steep increase in the GGM. We have compared and contrasted the GGM with a number of bipartite 
measures of quantum correlation. The GGM appears to have an advantage in detecting the quantum critical points,
particularly in the 2D frustrated quantum many-body systems.

\vskip 20pt
\noindent {\bf Acknowledgments}
\vskip 10pt

\noindent We thank Debasis Sadhukhan, Sudipto Singha Roy, and Titas Chanda for critical comments. 
R.P. acknowledges support from the Department of Science and Technology, Government of India, in the form of an INSPIRE faculty 
scheme at the Harish-Chandra Research Institute (HRI), India. We acknowledge computations performed at the cluster computing facility in HRI.
 We thank Indrani Bose and Amit Kumar Pal for useful discussions.
 
\vskip 20pt



\end{document}